\documentclass[12pt]{article}
\usepackage{graphicx}
\begin{document}
{\bf Interdisciplinary Monte Carlo Simulations}
\bigskip

D. Stauffer
\bigskip

Institute for Theoretical Physics, Cologne University, D-50923 Euroland
\bigskip

{\small Biological, linguistic, sociological and economical applications of
statistical physics are reviewed here. They have been made on a
variety of computers over a dozen years, not only at the NIC computers.
A longer description can be found in \cite{newbook}, an emphasis on
teaching in \cite{europe}.}

\section{Introduction}

The Monte Carlo methods invented for physics problems half a century
ago were later also applied to fields outside
of physics, like economy \cite{stigler}, biology \cite{kauffman}, or sociology
\cite{schelling}. Instead of atoms one simulates animals, including
people. These physics methods are often called ``independent agents'' when 
applied outside physics, to distinguish them from ``representative
agent'' approximations and other mean field theories. ``Emergence'' in
these fields is what physicists call
self-organization, that means systems of many simple particles showing complex
behaviour (like freezing or evaporating) which is not evident from the 
single-particle properties. 

The three people cited in Refs.3-5 were not physicists; two got the
economics Nobel prize. But also physicists have entered these fields 
intensively in the last years (and much earlier for biology; see Erwin 
Schr\"odinger's question: What is life?).
The German Physical Society has since several years a working
group on socio-economic problems, started by Frank Schweitzer.  And
our university just got approved a new Special Research Grant (SFB)
where geneticists and theoretical physicists are supposed to work together.
The NIC Research Group in J\"ulich is an earlier physics-biology example.

An important difference between physics and applications outside physics is the
thermodynamic limit. A glass of Cologne beer has about $10^{25}$ water molecules,
which is close enough to infinity for physicists. Economists, in contrast,
are less interested in stock markets with $10^{25}$ traders. Thus finite-size
effects, which often are a nuisance in Statistical Physics simulations, may be
just what we need outside of physics. 

Of this large area of computer simulations by physicists for fields
outside physics I now select: population genetics, language competition,
opinion dynamics, and market fluctuations, mostly following 
\cite{newbook,europe}.

\begin{figure}[hbt]
\begin{center}
\includegraphics[angle=-90,scale=0.3]{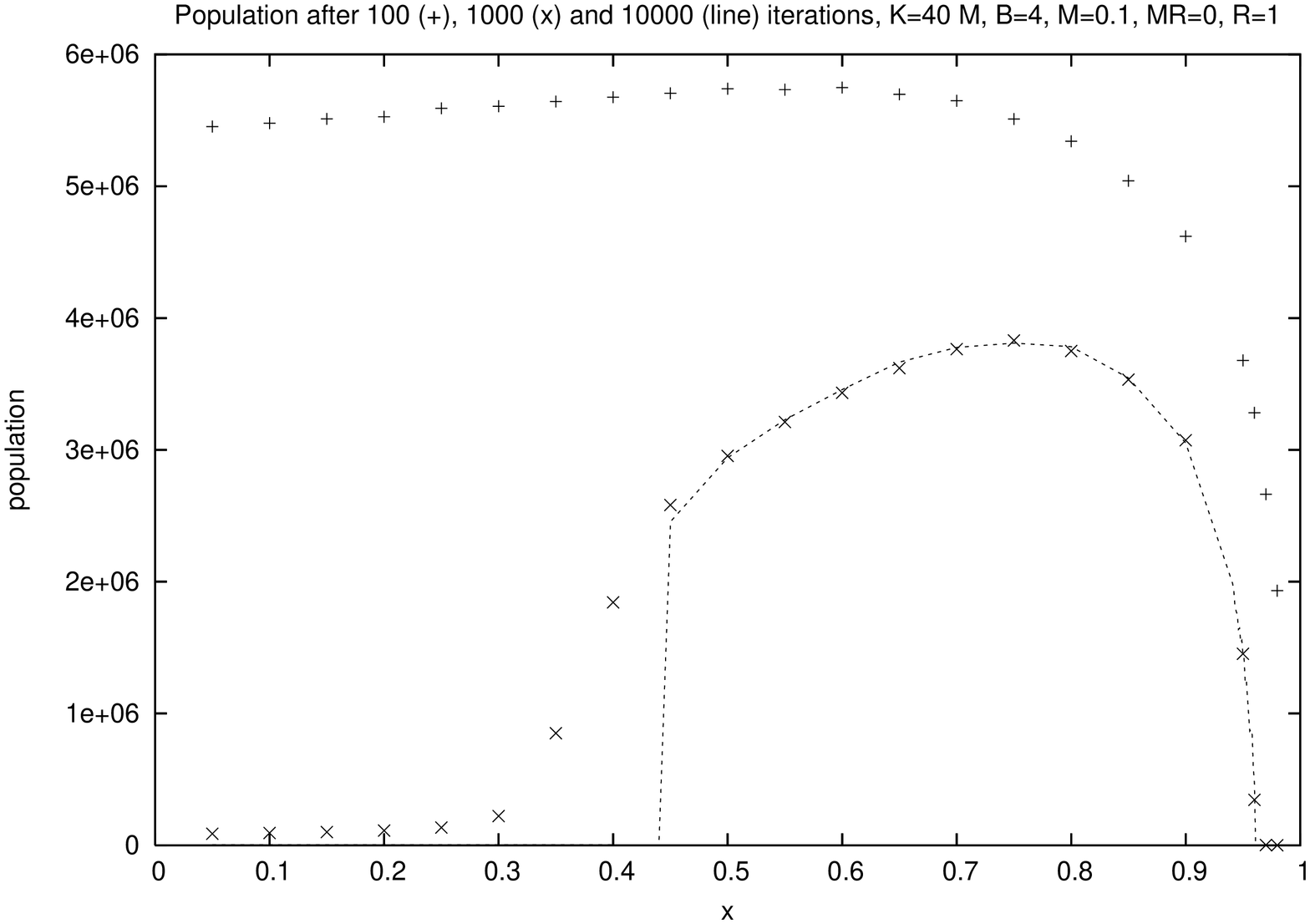}
\includegraphics[angle=-90,scale=0.3]{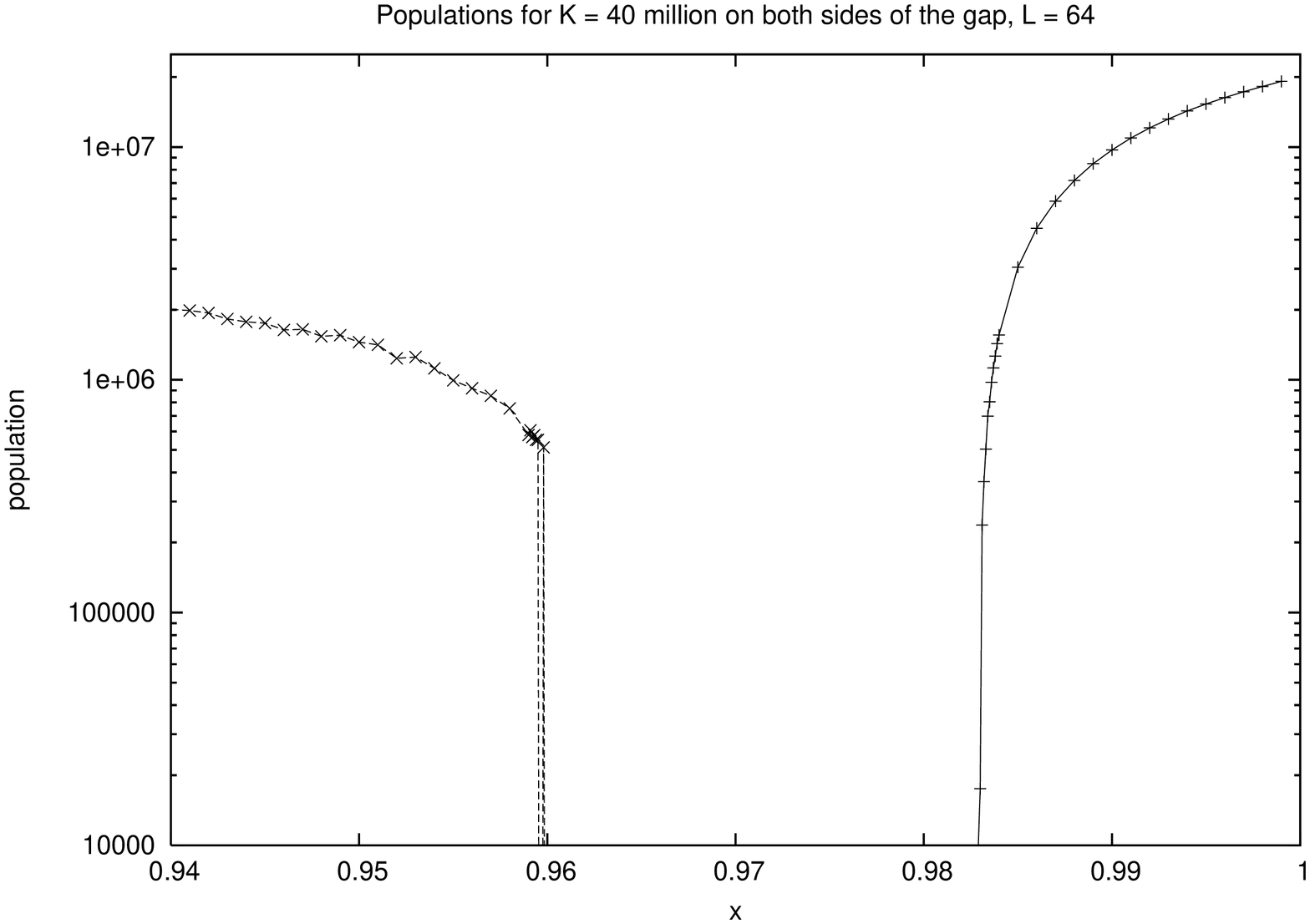}
\end{center}
\caption{$M=0.1, \; M_R=0, \; R=1, \; B=4, \; L=64$. Top part: First and second 
  phase transition, for various observation times; the third one
  at $x = 0.983$ is not shown for clarity. Bottom part: Expanded 
  semilogarithmic view of second and third phase transition.  
}
\end{figure}

\begin{figure}[hbt]
\begin{center}
\includegraphics[angle=-90,scale=0.5]{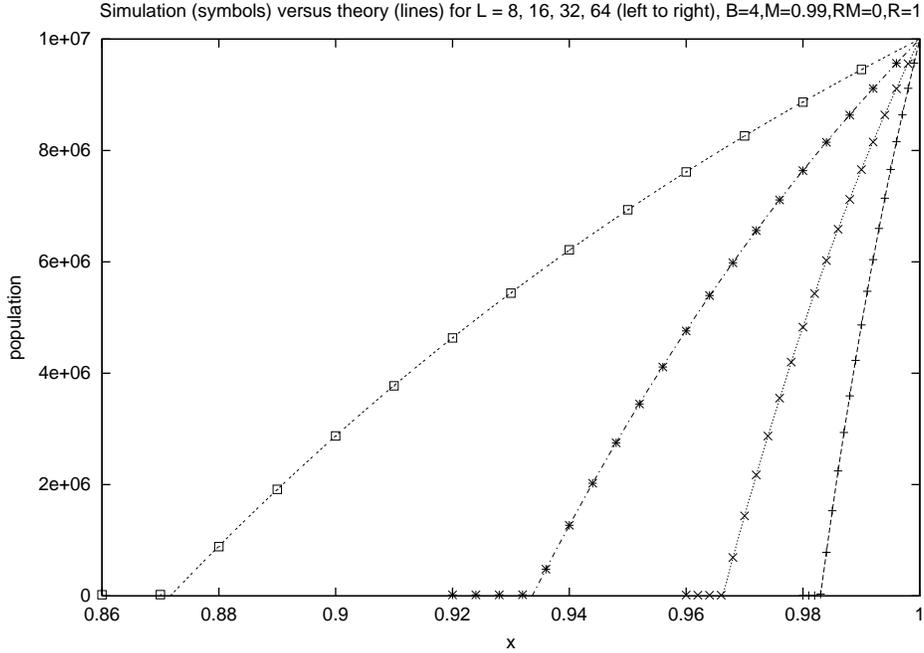}
\end{center}
\caption{Simulation (symbols) versus theory (lines) for the large-$x$ region
at $L=8$, 16, 32 and 64 (from left to right).
}
\end{figure}

\section{Population Genetics}
Darwinian Evolution is similar to thermal physics in that two effects compete:
Mother Nature wants to select the fittest and to minimize energy; but more
or less random accidents (mutations in biology, thermal noise or entropy in 
statistical physics) lead to deviations from ideality, like biological ageing
or minimization of the free energy. The following example is ongoing
work together with Cebrat, P\c ekalski, Moss de Oliveira and de Oliveira and
can be regarded as an improved Eigen quasispecies model.

Each individual in the population has a genome, which consists of two 
bit-strings inherited from the mother and the father, respectively. 
Each bit-string has $L$ bits with $L$ = 8,\,16,\,32,\,64, as is convenient
for Fortran words (byte to integer*8). A bit set to one means a bad
mutation in the DNA, while a zero bit is a healthy gene. All mutations
are assumed to be recessive, that means they diminish the survival
probability by a factor $x < 1$ if and only if both the paternal and
the maternal bit-string have their corresponding bits mutated. At
reproduction, the bit-strings in both the father and the mother are
mutated with probability $M$ at a randomly selected position; then
with probability $R$ they undergo a crossover (recombination) at some
randomly selected position (like in genetic algorithms); then the bits
neighbouring the crossover point are mutated with probability $M_R$;
and finally one bit-string of the mother and one of the father give
one child genome, with $B$ such births per iteration and per
female. (The mother selects the father at random.) Mutation attempts
for an already mutated bit leave this bit unchanged. 

At each iteration the genetic survival probability is $x^n$ where $n$ is the
number of active mutations (bit-pairs set to 1) and $x$ an input
parameter. To account for limitations in space and food, as well as
for infections from other individuals, additional Verhulst death
probabilities proportional to the current number of individuals are
applied to both the newborns and at each iteration to the adults.  

For very small $x$, only mutation-free individuals survive: $n=0$. With 
growing $x$ the survival chances grow, but so does the mutation load $<n>$ 
which in turn reduces the survival chances. As a result, for $L = 64$
three different phase transitions can be found in Fig.1: For $0 < x
< 0.45$ the population dies out; for $0.45 < x < 0.96$ it survives;
for $0.96 < x < 0.98$ it dies out again, and for $0.98 < x < 1$ it
survives again. The transitions at 0.45 and 0.96 seem to be
first-order (jump in population and load) while the one at 0.98 is
second-order (continuous). For $x > 0.98$ all bits of both bit-strings
are mutated to one, which allows a simple scaling prediction of the
population for general $L$ in agreement with the simulations: Results
depend on $x^L$ as seen in Fig.2. For example, the critical point at birth 
rate $B$ is at $x = (1 + B/2)^{-1/L}$.

Real animals get old with increasing age, and that can be simulated with
similar techniques. The more complicated Penna bit-string model \cite{penna}
simulates the ageing of
individuals and agrees well with the empirical Gompertz law of 1825,
that the mortality of adult humans increases exponentially with age 
\cite{newbook}.

\section{Language Competition}

Every ten days on average one human language dies out. Simulations of
the bit-string Schulze model are very similar to the above population
genetics, with random mutations, transfer of language bits from one language to
another, and flight from small to large languages \cite{schulzestauffer}. 
The alternative Viviane model \cite{viviane} simplifies mutation and
flight from small to large languages into one process, and ignores
transfer. It gives in Fig.3 a wide range of language 

\begin{figure}[hbt]
\begin{center}
\includegraphics[angle=-90,scale=0.30]{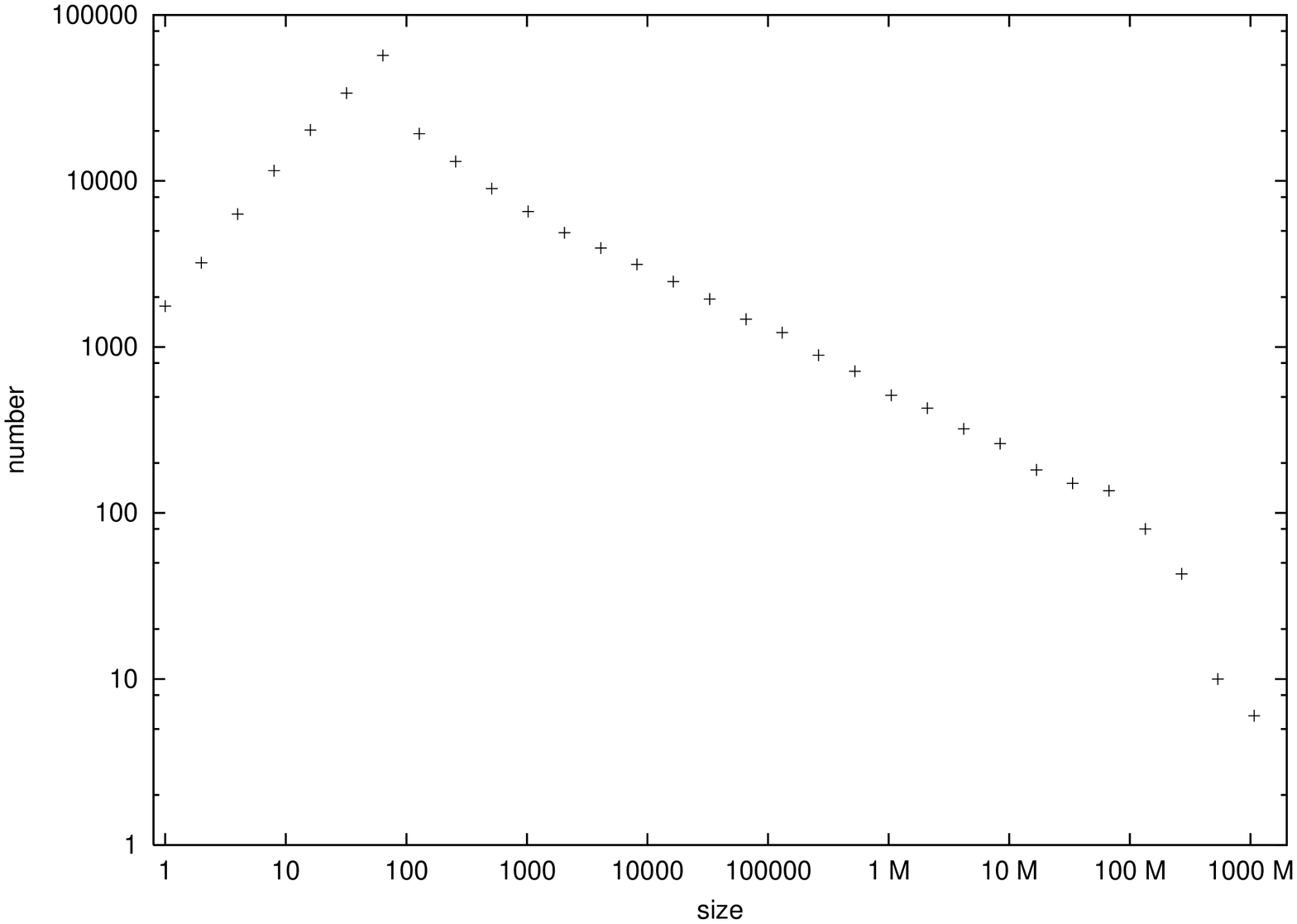}
\includegraphics[angle=-90,scale=0.30]{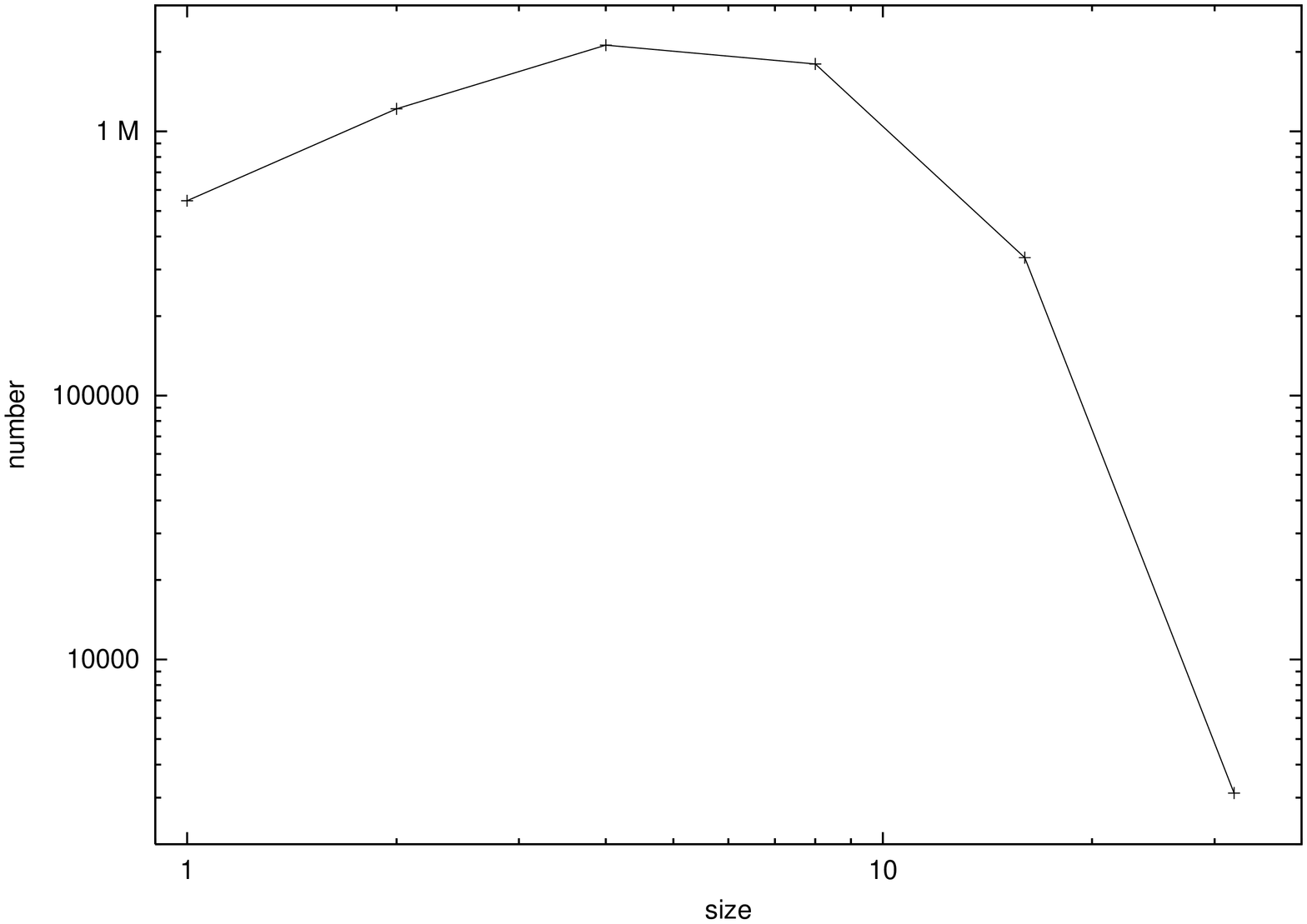}
\includegraphics[angle=-90,scale=0.30]{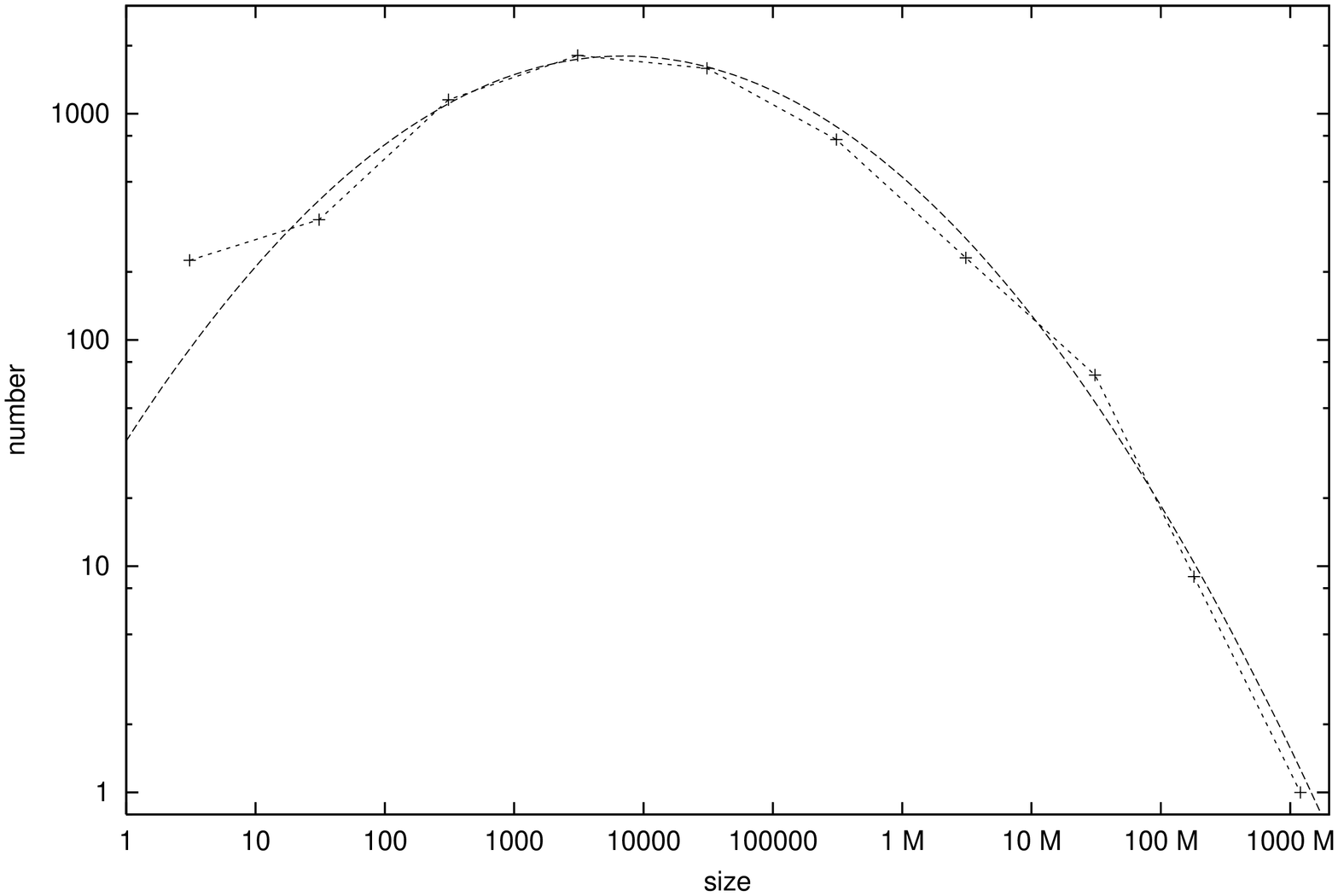}
\end{center}
\caption{Distribution of language sizes in Viviane model \cite{viviane} (top),
in Schulze model \cite{schulzestauffer} (middle) and reality \cite{sutherland}
(bottom). The curve in the bottom part is a log-normal fit.
}
\end{figure}
\clearpage

\noindent
sizes, i.e. of
the number of people speaking one language, from dying languages with
only one speaker, to Chinese with $10^9$ speakers. The Schulze model gives
a more realistic nearly log-normal shape for this distributions, but not the
wide range of language sizes. Both the proper shape and the large size range of 
reality (bottom part of Fig.3) might come from non-equilibrium statistics.

In the last version of the Schulze model, each language (better interpretation:
its grammar) is characterized by $F$ features each of which can adopt one of 
$Q$ different integer values $1,2,...Q$. Each site of a large square lattice is
occupied by a person speaking one language. At each iteration, each feature
of each person is mutated with probability $p$. This mutation is random with
probability $1-q$ while with probability $q$ the corresponding feature from 
one of the four lattice neighbours is adopted. Also, at each iteration, each
person independently, with
a probability proportional to $1-x^2$ abandons the whole language and adopts
the language of one randomly selected person in the population.

In the last version of the Viviane model, each lattice site is either empty of
carries a population with a size randomly fixed between 1 and, say, like 127.
Initially one lattice site is occupied and all others are empty. Then at each
time step one empty neighbour of an occupied site is occupied with a probability
proportional to the number of people which can live there. Then this new site
adopts the language of one of its four lattice neighbours, with a probability 
proportional to the size of the language spoken at that neighbour site. However,
this adopted language is mutated to a new language with probability inversely
proportional to 
the new size of the adopted language. (This denominator is not
allowed to exceed a maximum, set randomly between 1 and, say, 2048.)
The whole process ends once the last lattice site has become occupied.

\section{Opinion Dynamics}

Can a single person make a difference in public life? In chaos theory
we ask whether a single butterfly in Brazil can influence a hurrican
in the Caribbean. Kauffman \cite{kauffman} asked the analogous question whether
a single biological mutation has a minor effect or disturbs the whole
genetic network \cite{kauffman}. Physicists call this damage spreading
and ask, for example, how the evolution of an Ising model is changed
if one single spin is flipped and otherwise the system, including the
random numbers to simulate it, remains unperturbed. This question was
discussed \cite{fortunato,newbook} for three models: The opportunists
of Krause and Hegselmann \cite{krause}, the negotiators of Deffuant
et al \cite{deffuant}, and the missionaries of Sznajd \cite{sznajd}. 

The opportunists take as their new opinion the average opinion of the large 
population to which they belong, except that they ignore those who differ too
much from their own opinion. Also the negotiators ignore opinions which differ
too much from their own; otherwise a randomly selected pair gets closer in their
two opinions without necessarily agreeing fully. A randomly selected pair of 
missionaries, neighbouring on a lattice or network, convinces its neighbours if
and only if the two people in the pair have the same opinion. Simulations show 
that the opinion change of a single person may influence the whole population 
for suitable parameters \cite{fortunato,newbook}. 

For the missionaries on a scale-free network, simulations agreed nicely with 
election results in Brazil, apart from fitted scale factors, Fig.4.

\begin{figure}[hbt]
\begin{center}
\includegraphics[angle=-90,scale=0.5]{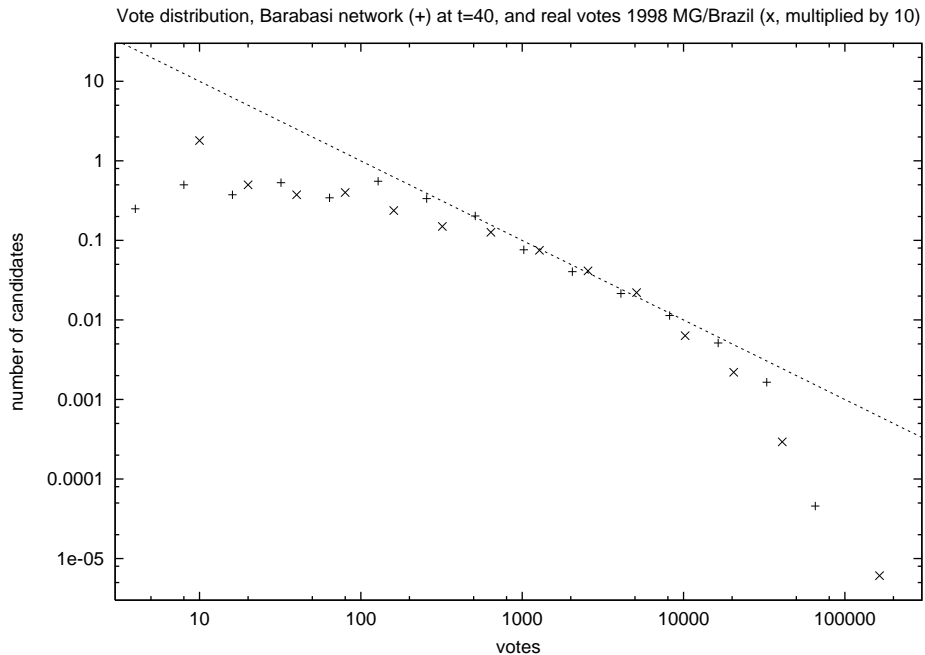}
\end{center}
\caption{Distribution of the number of candidates getting a certain number of
votes in simulations \cite{kertesz} and in elections in Minas Gerais, Brazil.
}
\end{figure}

\begin{figure}[hbt]
\begin{center}
\includegraphics[angle=-90,scale=0.5]{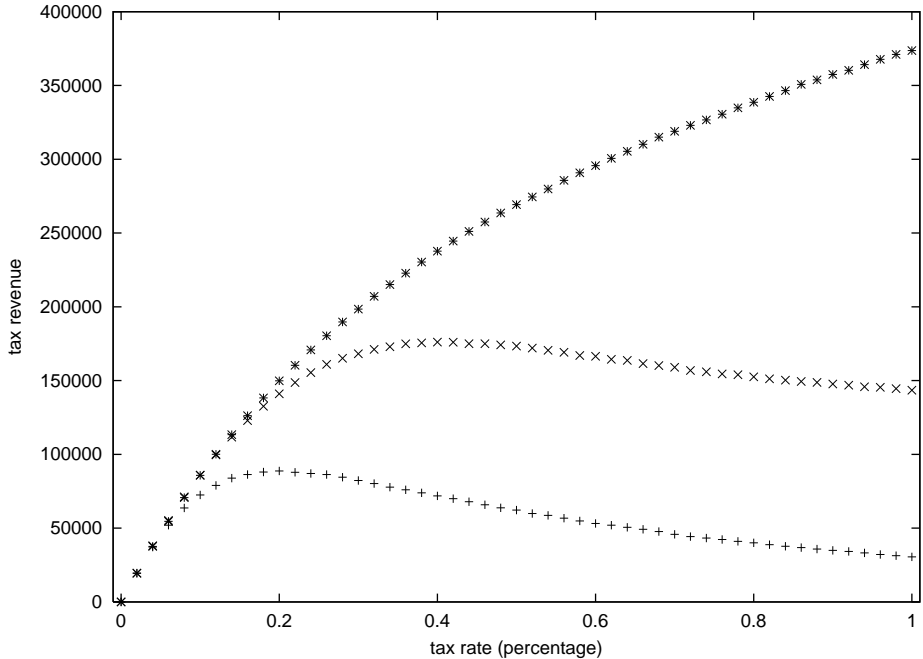}
\end{center}
\caption{Tax revenue for the government versus percentage of Tobin tax  to be
paid for each transaction, in various versions of the Cont-Bouchaud model
\cite{ehrenstein}. 
}
\end{figure}

\section{Market Fluctuations}

How can we get rich fast by speculating on the stock market?
This writer earned about one Heugel (a local currency unit of about $10^4$ Euro)
by believing some theory for the Tokyo stock market \cite{sornette}. Details,
of course, are given out only for more JUMP time. Instead this section 
summarizes the Cont-Bouchaud model of stock market fluctuations \cite{cont},
because it is closest to the pre-existing physics model of percolation. 

Each site of a large square lattice is either occupied by an investor (with
probability $p$), or empty with probability $1-p$. Sets of occupied neighbours
are called clusters and are identified with groups of investors which act (buy 
or sell) together. At each iteration a cluster either buys (with probability 
$a$), sells (also with probability $a$) or sleeps (with probability $1-2a$). The 
traded amount is proportional to the number of investors in the trading 
cluster. The difference between supply and demand drives the market values up 
and down. This basic model gives on average: i) as many ups as downs on the 
market; ii) a power-law decay (``fat tail'') for the probability to have a 
large price change, and with modifications also: iii) volatility
clustering (markets have turbulent and calm times), iv) effective 
multi-fractality, v) sharp peaks and flat valleys for the prices, but no 
prediction on how the market will move tomorrow. 

Apart from these nice basic properties also practical applications were made
\cite{ehrenstein}: Does a small ``Tobin'' tax of a few tenths of a percent
on all transactions reduce
fluctuations and earn tax revenue without killing the whole market? It does, but
apart from more government control over individuals there is another danger 
which can be simulated: If the tax revenue increases with increasing tax rate, 
then governments will be tempted to 
increase this tax again and again (as Germans just saw in fall 2005 and German
student may observe in future tuition hikes.) Much better is a maximum of tax
revenue at some moderate tax rate; then the government should settle on this 
moderate tax rate, provided it regards the simulations as reliable. Fig.5 shows that in this model such a 
desirable maximum exists for some parameters but not for all. Another 
application is the confirmation that halting the trade when excessive price 
changes are observed indeed helps to calm the market.

\section{Discussion}

Interdisciplinary applications of physics methods are no longer as exotic as 
they were years ago; biologists and economists have started to publish papers
together with computational physicists on these non-physics fields.

Thanks for S. Cebrat, P.M.C. de Oliveira and S. Moss de Oliveira for comments on
the manuscript.

\end{document}